# Evaluation of Battery Energy Storage System to Provide Virtual Transmission Service

Qiushi Wang and Xingpeng Li

*University of Houston, 4222 Martin Luther King Blvd, Houston, 77204, Texas, USA*

---

## Abstract

An immediate need in the transmission system is to find alternative solutions that improve system operation and defer the need for new transmission lines. This study comprehensively evaluates the performance and economic benefits of short-term operation of using battery energy storage systems (BESS) as virtual transmission (VT) to promote power transfer across distant regions. Specifically, this work implements various day-ahead energy scheduling models to analyze the impact of VT on system operation cost, network congestion, model computational time, and market performance. The performance of VT is compared with three alternative network congestion mitigation methods, including building new high-voltage physical transmission lines, cost-driven battery energy storage systems, and network reconfiguration, as well as combinations of two of the aforementioned methods. The benchmark day-ahead scheduling model is a traditional security-constrained unit commitment model without system upgrades or other network congestion mitigation. Numerical simulations conducted on the IEEE 24-bus system demonstrate that VT provides a 14% operational cost reduction and 34% congested line relief compared to the base case. Compared to other examined schemes, VT is the only one comparable to physical transmission lines that can provide satisfying congestion relief and operation cost reduction without significantly sacrificing computing time and load payment.

---

## NOMENCLATURE

| | |
|---|---|
| $g$ | Transmission element (line or transformer) index. |
| $k$ | Transmission element (line or transformer) index. |
| $n$ | Bus index. |
| $s$ | Solar generation index. |
| $w$ | Wind generation index. |
| $e$ | Battery index. |
| $vt$ | Virtual transmission line index. |
| $t$ | Time period index. |
| $G(n)$ | Set of generators at bus *n*. |
| $K$ | Set of all transmission elements. |
| $N$ | Set of all buses. |
| $ES(n)$ | Set of all battery storage systems at bus *n*. |
| $ES(vt)$ | Set of all battery-based virtual transmission systems. |
| $S(n)$ | Set of all solar generators at bus *n*. |
| $W(n)$ | Set of all wind generators at bus *n*. |
| $K(n-)$ | Set of branches with bus n as the to-bus. |
| $K(n+)$ | Set of branches with bus n as the from-bus. |
| $X_k$ | The reactance of transmission element. |
| $C_g$ | Linear cost for generator *g*. |
| $C_g^{NL}$ | No-load cost for generator g. |
| $C_g^{SU}$ | The start-up cost for generator g. |
| $C_s$ | The penalty cost for solar curtailment |
| $C_w$ | The penalty cost for wind curtailment |
| $d_{nt}$ | Predicted load demand of bus n in the time period t. |
| $BigM$ | A big real number. |
| $P_g^{min}$ | The minimum capacity of generator g. |
| $P_g^{max}$ | Maximum capacity of generator g. |
| $P_k^{max}$ | Emergency thermal line limit for line *k*. |
| $u_{gt}$ | Commitment status of unit *g* in the time period *t*. |
| $v_{gt}$ | Start-up variable of generator g in the time period *t*. |
| $\theta_{kt}$ | Phase angle difference between from-end and to-end of line k in the time period *t*. |
| $P_{gt}$ | The output of generator g in the time period *t*. |
| $P_{kt}$ | Flow in line k in the time period *t*. |
| $P_{st}$ | The output of solar generators in the time period *t*. |
| $u_{st}$ | Commitment status of solar unit s in the time period *t*. |
| $P_{wt}$ | The output of wind generators in the time period *t*. |
| $u_{wt}$ | Commitment status of wind unit s in the time period *t*. |
| $P_{st}^{C}$ | The curtailed solar generation in the time period t |
| $P_{wt}^{C}$ | The curtailed wind generation in the time period t |
| $P_{et}^{c}$ | Charging rate of battery e in the time period *t*. |
| $P_{et}^{d}$ | Discharging rate of battery e in the time period *t*. |
| $E_{et}$ | Energy storage energy level in the time period *t*. |
| $u_{et}^{c}$ | 1 indicates charging mode; otherwise, 0. |
| $u_{et}^{d}$ | 1 indicates discharging mode; otherwise, 0. |
| $E_{e}^{min}$ | Minimum energy storage energy level. |
| $E_{e}^{max}$ | Maximum energy storage energy level. |
| $P_{e}^{c,max}$ | Maximum energy storage charge rate. |
| $P_{e}^{d,max}$ | Maximum energy storage discharge rate. |
| $\eta_{e}^{c}$ | Charging efficiency. |
| $\eta_{d}^{d}$ | Discharging efficiency. |
| $\rho_{s}$ | The solar penalty coefficient |
| $\rho_{w}$ | The wind penalty coefficient |
| $\Delta T$ | Length of a time interval. |
| $J_{kt}$ | 1 indicates branch *k* is in the network in the time period *t*; otherwise, it is 0. |

## 1. Introduction

Although fast growth of solar and wind power substantially decarbonizes the electricity sector, a large portion of clean energy generation is expected to be frequently curtailed and thus wasted due to limited transmission capacity. Even with the current penetration level of renewable generation in many practical power grids, curtailment of clean energy generation is often observed. For example, in California Independent System Operator (ISO) territory, 187,000 MWh of wind and solar generation was curtailed in 2015. The curtailment amount increased by a factor of 8.5 to 1,587,000 MWh in 2020 [1]- [2].

The United States Department of Energy published the 2023 National Transmission Needs Study. [3] In the report, the team concluded that an immediate need for updated or new transmission infrastructure is required by 2030 to meet the load growth and clean energy penetration. However, the overall transmission investment has decreased over the past decades, while the average timeline for building a new high-voltage transmission line is approximately ten years. To bridge the gap between short-term transmission needs and long-term transmission planning and deployment, non-wire alternatives that can help alleviate transmission network congestion and reduce renewable generation curtailment are investigated and compared to traditional wired solutions in this paper.

One of the non-wire solutions to relieve congestion and reduce renewable curtailment is a large-scale battery energy storage system (BESS). BESS can help reshape the load profile and thus impact the power flow in the adjacent area. Alberto and Steven provided a brief overview of existing energy storage technologies and their applications in [4]. They also illustrated using battery storage to increase transmission capability by relaxing *N*-1 contingency conditions in thermally constrained networks of high-voltage transmission lines. Article [5] and [6] further developed the economic dispatch algorithm to enable merchant storage facilities to compete in an electricity market to provide transmission congestion relief services.

A few BESS studies on real-world power systems in the literature [7] and [8] concurred with the research conclusion that adding BESS can help reduce congestion and offset transmission needs. Pacific Northwest National Laboratory examined the technical and financial feasibility of using a BESS and a combustion turbine generator (CTG) to defer the investment in a third transmission cable for Nantucket Island [9]. The assessment results showed that the benefits of BESS plus CTG operations with minimal, low-cost distribution upgrades outweighed constructing a third transmission line.

Besides the stationary battery storage system, prior efforts in the literature also evaluated the possibility of mobile energy storage systems. There is a feasible solution of the integrated optimization model for distribution planning problems, which uses temporary, transportable energy storage to reduce or defer the distribution network expansion. [10]. The integrated planning strategy's application to the transmission network was further extended. [11] [12] The researchers presented different algorithms to minimize the cost of a combination of transmission lines and battery-based energy storage units (BESS), which include stationary and mobile storage.

Most BESS-related research and applications demonstrate the benefit of a single BESS application to improve local system performance. [13] [14] With continuous technological advancement, BESS has become more cost-effective, and its size and duration have increased. [15] [16] As a result, BESS may provide more benefits in its existing applications and gain the potential to support new applications that need to be explored. However, very few BESS-related research studies discuss the impact of additional Virtual Transmission (VT) constraints.

BESS-based VT is another non-wire alternative to traditional transmission lines. It can help increase the power transfer capability and relieve the load flow burden of the existing power network [17]. Like the standalone BESS, BESS-based VT has a much shorter implementation time than the physical transmission line. Unlike the standalone BESS, BESS-based VT emphasizes coordination between the two BESSs to mimic the operation of the Physical Transmission (PT) line.

Nguyen demonstrated the VT concept in a two-machine network, which uses BESSs at the two ends of a line to mimic a power line [18]. The objective of VT in [18] is to increase revenue for generators in the region during congested and non-congested times. The coordination strategy of the two BESSs is activated to provide virtual transmission capacity when the targeted line is congested, while it is disabled during non-congested times. Nanou's team adopted Nguyen's congestion condition-based coordination strategy and evaluated this strategy with the objective of minimizing operating costs [19]. Nguyen's study and Nanou's study show that VT's congestion condition-based coordination strategy can help reduce LMP and increase virtual line capacity. However, the performance of the VT under this coordination strategy designates the supply and demand sides of the VT, which prevents the VT from functioning well if the direction of power flow changes.

Another objective of the VT application is to minimize the total relative congestion level. In [20], Lindner's team evaluated the congestion management (CM) performance of grid operator-owned VT lines where there is no interface with the energy market. Lindner assumes that the VT coordinates so that the BESSs won't affect the power balance of the existing network at any time. When the BESS is used in preventive CM mode, it is referred to as VT, while it is referred to as grid booster (GB) when used for curative CM. The researchers found that using GB as a curative CM is more effective than VT as a preventive CM for the battery size and location in the test network.

Although ongoing pilot VT projects are happening globally [21], minimal research has been done to understand how VT schemes would behave in a non-conditional-based coordination strategy and a deregulated market environment. In addition, it is also crucial to investigate how well VT performs compared to other CM schemes and how well VT coordinates with other CM schemes.

Network reconfiguration (NR) has been demonstrated to be a low-cost but very effective CM strategy in both transmission and distribution systems [22] [23] [24]. NR can relieve network congestion as a preventive control scheme in the pre-contingency situation [25] and as a corrective control scheme in the post-contingency situation [26] [27]. As a congestion relief strategy, NR achieves substantial system cost savings and reduces significant renewable generation curtailment [28] [29].

To bridge the research gaps and address the question of how VT behaves compared with other non-wire alternatives, this paper will investigate the effectiveness of VT as a non-wire transmission capacity expansion solution in the application of day-ahead operational planning that solves the security-constrained unit commitment (SCUC) problem. Its performance will be evaluated and compared to other options, including new high-voltage physical transmission (PT) lines and NR. In addition, this paper will also combine multiple CM options to achieve better grid performance.

Particularly, this paper will implement seven different SCUC optimization models to evaluate various congestion mitigation schemes for day-ahead generation scheduling. These seven SCUC models are explained as follows: (1) a traditional benchmark SCUC, (2) an enhanced SCUC with a new physical transmission (SCUC-PT), (3) an enhanced SCUC with BESS (SCUC-BESS), (4) an enhanced SCUC with non-simultaneous charging and discharging constraints on BESS as VT (SCUC-VT), (5) an enhanced SCUC with NR (SCUC-NR), (6) an enhanced SCUC with both VT and NR (SCUC-VT-NR), and (7) an enhanced SCUC with BESS and NR (SCUC-BESS-NR).

The remainder of this paper is organized as follows. The formulations for various SCUC models of interest are described in Section II. The test case and simulation results are presented in Section III. Finally, Section IV concludes this paper and presents potential future work.

## 2. Methods

The day-ahead energy scheduling of the power system is determined by solving SCUC. This section presents the formulations used by a traditional SCUC model as a benchmark and various enhanced SCUC models with congestion mitigation strategies. It also explains the metrics used to analyze the impacts of different CM strategies on the wholesale power markets.

### *2.1 Traditional SCUC*

SCUC minimizes the total cost of generations over multiple time periods while maintaining the solution physically feasible for each period. A widely used formulation for a traditional SCUC model is presented as follows.

Objective:

$$minimize \sum_{g \in G} \sum_{t \in T} \left( c_g P_{gt} + c_g^{NL} * u_{gt} + c_g^{SU} * v_{gt} \right) \tag{1}$$

Constraints:

$$u_{gt} \in \{0, 1\}, \forall g, t \tag{2}$$

$$v_{gt} \in \{0,1\}, \forall g, t \tag{3}$$

$$v_{gt} \geq u_{gt} - u_{g,t-1}, \forall g, t \tag{4}$$

$$P_g^{min} * u_{gt} \le P_{gt}, \forall g, t \tag{5}$$

$$P_{gt} \le P_g^{max} * u_{gt}, \forall g, t \tag{6}$$

$$P_{gt} - P_{g,t} \le R_g^{hr}, \forall g, t \tag{7}$$

$$P_{g,t-1} - P_{g,t} \le R_g^{hr}, \forall g, t \tag{8}$$

$$P_{kt} = \theta_{kt}/x_k, \forall k, t \tag{9}$$

$$-P_k^{max} \le P_{kt} \le P_k^{max}, \forall k, t \tag{10}$$

$$\begin{aligned} \sum_{g \in G(n)} P_{gt} + \sum_{k \in K(n-)} P_{kt} - \sum_{k \in K(n+)} P_{kt} \\ = d_{nt} - \sum_{s \in S(n)} P_{st}, \forall n, t \end{aligned} \tag{11}$$

The objective function (1) minimizes the system's total cost, including generator operation, start-up, and generators' no-load costs. Equations (2)-(11) are the constraints for the traditional SCUC optimization model. Binary variables $u_{gt}$ for generation commitment status and $v_{gt}$ for generator startup indicator are defined in (2) and (3), respectively. Constraint (4) defines the relation between $u_{gt}$ and $v_{gt}$. Generator output limits are enforced in (5)-(6). Generator ramping rate limits are respected in (7)-(8). The line power flow equation and thermal capacity limit are presented in (9) and (10), respectively. Constraint (11) guarantees the power balance will be met at each node in each time interval.

*2.2 Network Congestion Mitigation Solutions*

This sub-section will first present the formulations for several CM schemes, including BESS, VT, PT, and NR. The corresponding SCUC models are then summarized.

When BESS is present in the system to be scheduled along with generators, some existing constraints need to be updated to capture the impact of BESS. At the same time, new constraints are also required to represent BESS's unique characteristics in SCUC. BESS cannot be charged or discharged at the same time. Instead, the status of a BESS should be either charging, discharging, or idle, as represented in (12). When the BESS is in charging or discharging mode, the charging and discharging power rate must be within the maximum physical limits as enforced in (13)-(14). Constraint (15) sets the boundaries of BESS energy level. In (16), the BESS energy level calculation considers charging and discharging efficiencies. In a network with BESSs, (11) needs to be replaced by the updated nodal power balance constraint (17) due to BESS charging and discharging activities.

$$u_{et}^{c} + u_{et}^{d} \leq 1, \forall e, t \tag{12}$$

$$0 \leq P_{et}^{c} \leq P_{e}^{c,max} u_{et}^{c}, \forall e, t \tag{13}$$

$$0 \leq P_{et}^{d} \leq P_{e}^{d,max} u_{et}^{d}, \forall e, t \tag{14}$$

$$E_{e}^{min} \leq E_{et} \leq E_{e}^{max}, \forall e, t \tag{15}$$

$$E_{et} = E_{e,t-1} + \left(\eta_{e}^{c} P_{et}^{c} - P_{et}^{d} / \eta_{e}^{d}\right) \Delta T \tag{16}$$

$$\begin{aligned} \sum_{g \in G(n)} P_{gt} + \sum_{k \in K(n-)} P_{kt} - \sum_{k \in K(n+)} P_{kt} \\ = \sum_{e \in ES(n)} \left(P_{et}^{c} - P_{et}^{d}\right) + d_{nt} - \sum_{s \in S(n)} P_{st}, \forall n, t \end{aligned} \tag{17}$$

The significant difference between standalone BESS and BESS-based VT is the coordination between the two BESSs — the two BESSs operating under a virtual transmission line strategy coordinate to mimic the operation of a PT line. A transmission line absorbs the power from one end while injecting that power into the other end. Therefore, to ensure the behavior of BESS-based VT is consistent with a parallel PT line, constraints are needed to avoid BESSs on both sides of the transmission line charging simultaneously or discharging simultaneously.

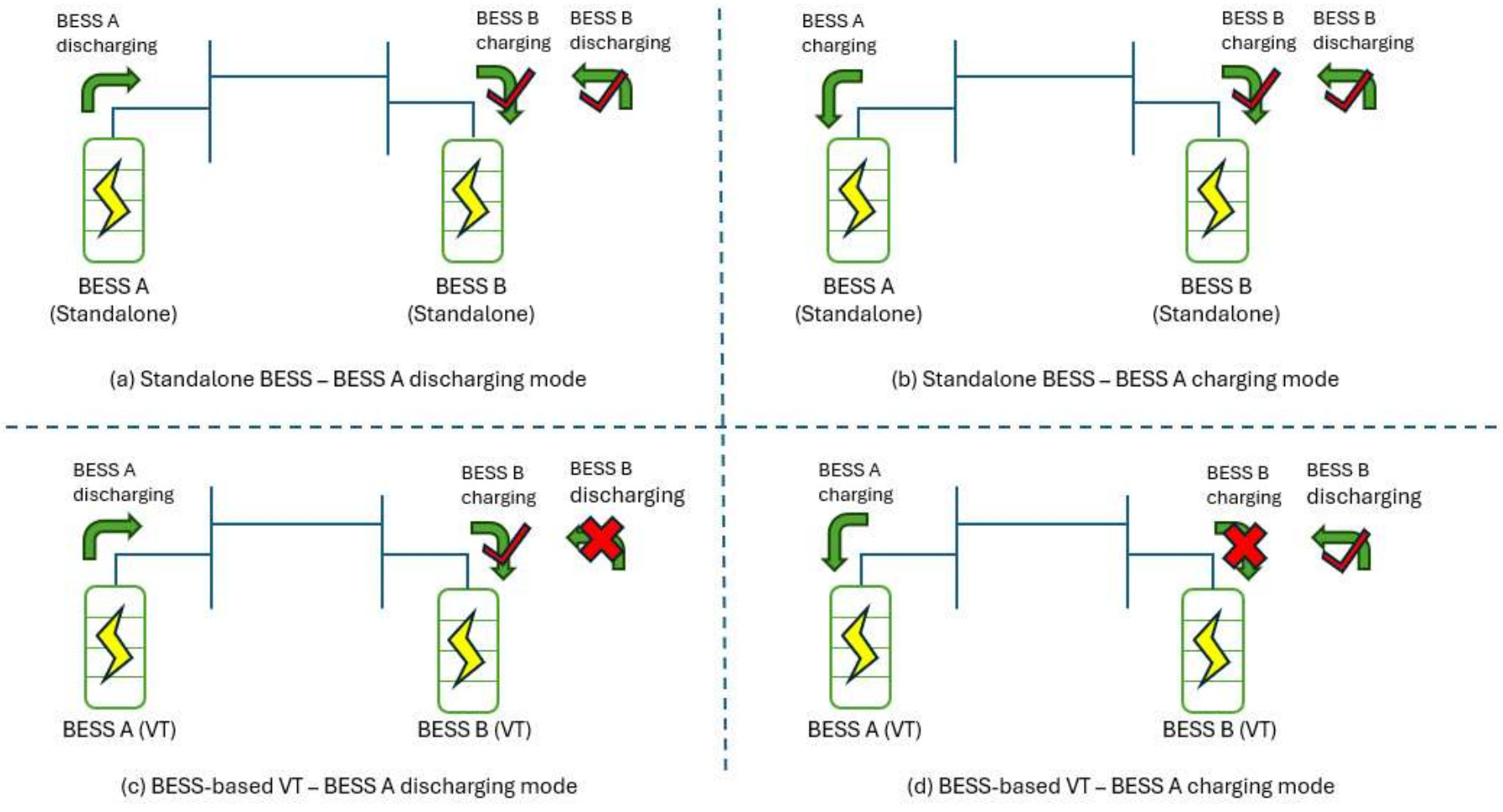

Figure 1 Illustration of differences on coordination strategy between standalone BESS and BESS-based VT

Figure *1* illustrates different coordination strategy between standalone BESS and BESS-based VT. Standalone BESS has no coordination to guide other standalone BESS's charging/discharging decisions. As shown in Figure 1 (a), When standalone BESS A discharges power to the grid, standalone BESS B can either discharge power to the grid or charge power from the grid. On the other hand, the two BESS operating in a virtual transmission line strategy coordinate with each other. As shown in Figure 1 (c), VT's BESS B is not allowed to be in the discharging mode if VT's BESS A is in the discharging mode. Instead, BESS B will be in the charging mode or idle when BESS A is in the discharging mode.

Equations (18)-(19) are constraints to limit the status of BESSs for VT lines, in which $ES(vt)$ is a set of two BESSs that are located at the two ends of a congested physical line respectively to ensure VT behavior for each $vt$.

$$\sum_{e \in ES(vt)} u_{et}^{c} \leq 1, \forall vt, t \tag{18}$$

$$\sum_{e \in ES(vt)} u_{et}^{d} \leq 1, \forall vt, t \tag{19}$$

Network reconfiguration that can leverage the flexibility in the transmission network is another effective method to help relieve line congestion. This study will also implement the NR scheme to evaluate the performance of stand-alone BESSs and VTs. The updated constraints when implementing NR in SCUC are listed as (20)-(23), replacing (9)-(10). The model also includes a constraint (21) to limit the number of line switching actions to at most one in a single time interval to avoid severe system stability risks.

$$J_{kt} \in \{0,1\}, \forall k, t \tag{20}$$

$$\sum_{k \in K} (1 - J_{kt}) \leq 1, \forall t \tag{21}$$

$$-BigM(1 - J_{kt}) \leq P_{kt} - \theta_{kt}/x_{k\ \leq BigM(1-J_{kt}), \forall k,t} \tag{22}$$

$$-J_{kt} P_{k}^{max} \leq P_{kt} \leq J_{kt} P_{k}^{max}, \forall k, t \tag{23}$$

Seven different SCUC optimization models for day-ahead generation scheduling are then formulated and implemented to evaluate various congestion mitigation schemes. They are explained in Table 1.

Table 1 Model Descriptions and Formulation Summary

| Models | Descriptions | Equations |
|---|---|---|
| SCUC | The traditional SCUC optimization model is a benchmark. | (1)-(11) |
| SCUC-PT | A new physical line is added to the system and the SCUC model. | (1)-(11) |
| SCUC-BESS | BESSs are added to the system and the SCUC model. | (1)-(10), (12)-(17) |
| SCUC-VT | VT operation constraints are added to the SCUC-BESS model. | (1)-(10), (12)-(19) |
| SCUC-NR | Network reconfiguration strategy is applied to the SCUC case. | (1)-(11), (20)-(23) |
| SCUC-BESS-NR | Network reconfiguration strategy is applied to the SCUC-BESS case. | (1)-(10), (12)-(17), (20)-(23) |
| SCUC-VT-NR | The network reconfiguration strategy is applied to the SCUC-VT case. | (1)-(10), (12)-(23) |

*2.3 Market Analysis Metrics*

Congestion management and transmission transfer capacity investment are essential to meet load growth and support clean energy penetration. It is also important to analyze the impact of those schemes on the wholesale power energy markets. In this paper, it is assumed that the power market follows a locational marginal price (LMP)-based market clearing mechanism that most US grid operators adopt.

LMP is the marginal cost of supplying one additional MW of power to a given location. It is dependent on not only the location but also the time. Mathematically, it is equal to the dual variable of the nodal power balance constraint. In addition to the total generation cost, another metric for evaluating the system efficiency and market performance is the load payment, which is defined as follows,

$$Load\ Payment = \sum_{n} \sum_{t} d_{nt} LMP_{nt} \tag{24}$$

## 3. Case Studies

All the aforementioned variations of SCUC models were implemented and tested on the IEEE 24-bus system, which has been modified to reflect the current trend of transforming coal power into more sustainable generation resources. The optimization problems were solved using the Gurobi solver in the Python-based Pyomo package. The Python scripts were run in the Anaconda Spyder environment with Python version 3.8.6 on Intel(R) Core(TM) i5-3570K CPU @ 3.40GHz 3.40 GHz computer system.

*3.1 Test Power System Case*

The IEEE 24-bus system was first developed in 1979 with a load model, generation system, and transmission network [30]. Since then, it has been widely used as a test system for transmission planning and reliability tests. The system has 24 buses with 38 connected elements at voltage levels of 230 kV and 138 kV. Figure 2 shows the configuration of the IEEE 24-bus system. The branch and conventional generation and load data can be found in [30]. The generator costs are adopted from [31].

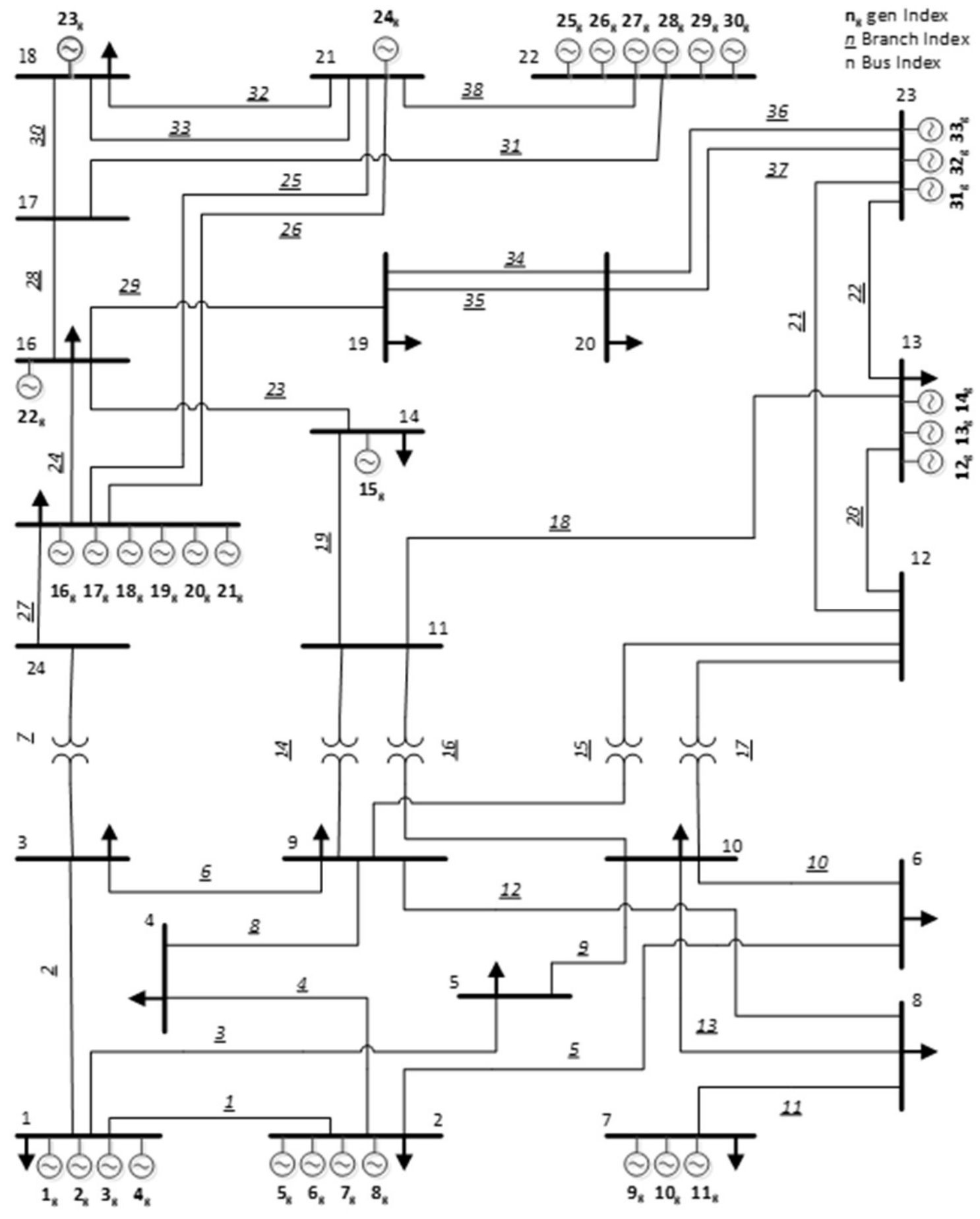

Figure 2. Network topology of the IEEE 24-bus system *[32]*.

Modifications to the generation are made to the case based on the assumption that the future power system will be free of coal-fired generation. All conventional coal-type generators on buses 2, 15, 16, and 23 are removed from the system. Instead, solar generators totaling 1,110 MW are added to buses 14, 15, and 16. In this paper, the solar deliveries are fixed at their maximum available power, as there will be no curtailment even when transmission lines are congested. The daily peak load is assumed to be 80% of the maximum load. The load profiled in the test cases uses summer weekday data in [30] as the hourly peak load in percent of daily peak. Unless specified otherwise, the load profile remains consistent across all scenarios in Section 3.

*3.2 Traditional SCUC Results*

Traditional SCUC optimization is run to evaluate the congestion level of the modified IEEE 24-bus network. Results show two lines with congestion: line 11 during evening hours and line 19 during peak sun hours. In addition to lines 11 and 19, line 29 gets stressed and operates above 70% of the line capacity between 10 a.m. and 4 p.m.

Line 11 is a generation tie line that connects generators 9-11 to the system through the point of connection bus 8. Line 11 hits its thermal limit mainly because generators 10 and 11 need to deliver power at their total capacity to meet the load profile when solar resources are unavailable after sunset. Since generation tie lines are usually designed to match the plant's maximum power output limit, upgrading line 11 for more transfer capacity against inter-area congestion is unnecessary.

On the other hand, line 19 is connected to one of the corridors between the 230 kV and 138 kV systems. The line gets congested when the system utilizes all the available solar power during peak sun hours. Applying new infrastructures to this line can help us better understand how BESS helps relieve congestion and reduce system costs in a meshed system. Therefore, line 19 is the targeted line to place new lines and BESSs. Additionally, adding new infrastructure to line 19 has the potential to help eliminate the congestion observed on line 11.

In the SCUC-PT line case, the new line is added in parallel to line 19 between buses 11 and 14 with identical line parameters as line 19. In BESS-related cases, both batteries on each side of line 19 are assumed to have the same size and technical specifications. The size of each BESS is 800 MWh with a maximum charging/discharging rate of 200 MW.

*3.3 Comparison of Simulation Results*

As the objective of the models is to minimize the total cost, this paper compares the system's economic performance under different CM schemes. In addition, the associated system congestion and computing time are analyzed and compared. These comparisons are summarized in Table 2.

Table 2 Transmission Facility Performance Comparison

| Model | Operation cost reduction | Average No. of congested lines per hour | Computing time (s) |
|---|---|---|---|
| SCUC | 0.00% | 0.38 | 6.5 |
| SCUC-PT | 11.71% | 0.13 | 5.4 |
| SCUC-BESS | 14.09% | 0.42 | 2.3 |
| SCUC-VT | 14.04% | 0.25 | 2.3 |
| SCUC-NR | 6.38% | 0.38 | 268.1 |
| SCUC-BESS-NR | 15.26% | 0.38 | 1446.2 |
| SCUC-VT-NR | 15.16% | 0.42 | 10300.4 |

The "operation cost reduction" column provides an overview of the percentage decrease in the daily total system operation cost compared to the SCUC base case. The study results show that all examined CM schemes can achieve cost reduction.

The "average No. of congested lines per hour" column describes the overall congestion status of the 24-bus system over the 24 hours. The higher the average number of congested lines per hour in the system, the more line(s) or the more hour(s) the line(s) are congested. Among all the CM schemes, only the PT and VT schemes provide Pareto improvement solutions and can well balance between system congestion relief and cost reduction. Although both BESS and VT schemes are implemented by adding two identical batteries on both sides of the transmission line, the standalone BESSs without VT constraints did not relieve line congestion but led to more network congestion. With VT constraints, the BESSs on both sides of the congested line can provide more virtual power transfer capability and implicitly mitigate network congestion. Subsection III D below provides a detailed analysis of the cause of different battery behaviors under BESS and VT schemes.

The NR is the only scheme that does not require additional capital investment costs and can substantially reduce the operation cost. However, one serious concern when using BESS/VT with NR in SCUC problems is the computational complexity since such scheme combinations would result in a much longer solving time than other schemes in an order related to the number of branches.

*3.4 BESS Operation Analysis*

The main difference between BESS and VT schemes is the restriction on battery charging and discharging status. The BESS scheme allows batteries to charge or discharge freely within their energy level limits. In contrast, the VT scheme prohibits the two batteries on each side of the transmission line from charging simultaneously or discharging simultaneously. The VT operation constraint leads to a different optimal battery operation solution for the testing system. Figure 3 and Figure 4 show the energy exchange profiles of the batteries on bus 11 and bus 14, respectively.

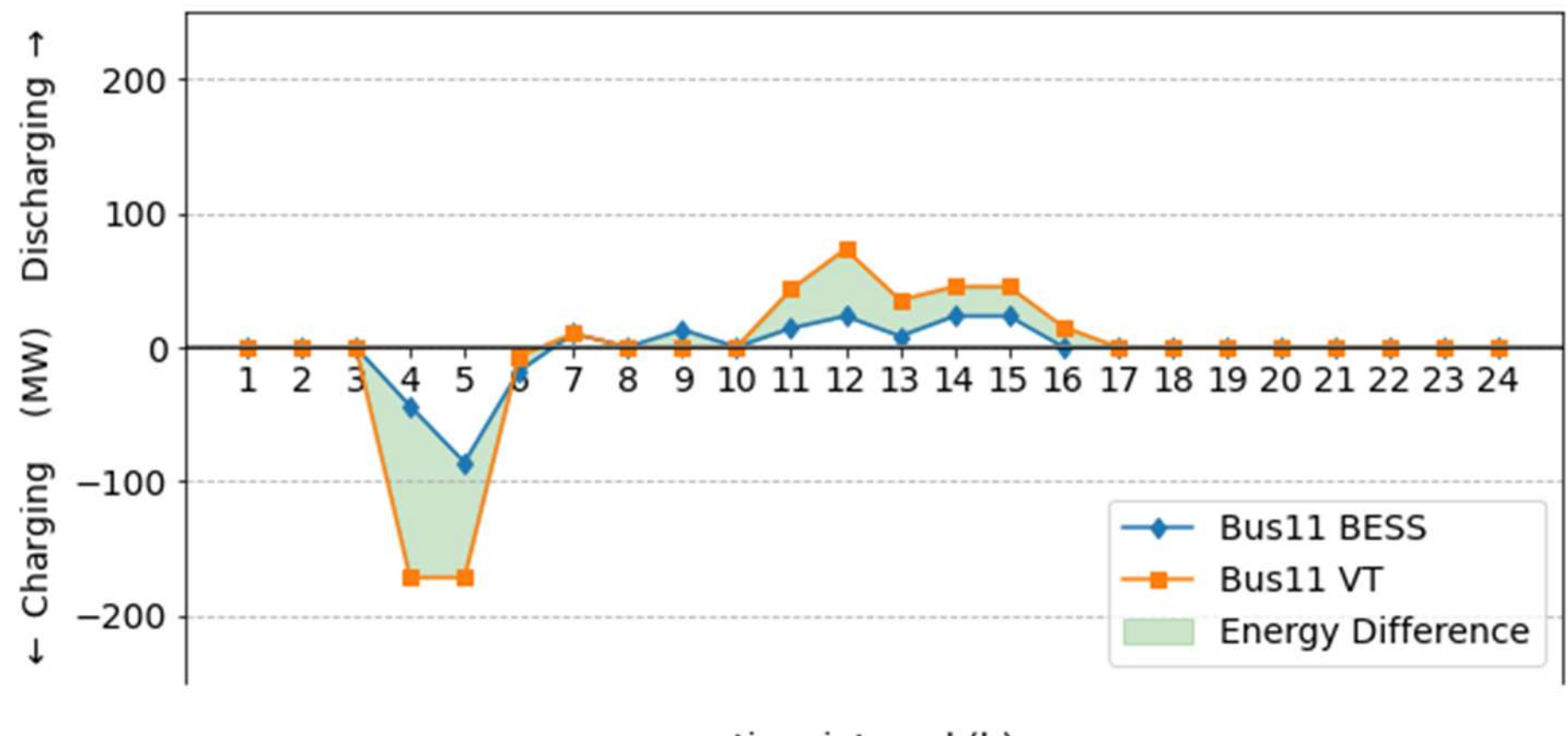

Figure 3. Charging and discharging profile of the battery at bus 11 under different CM strategies.

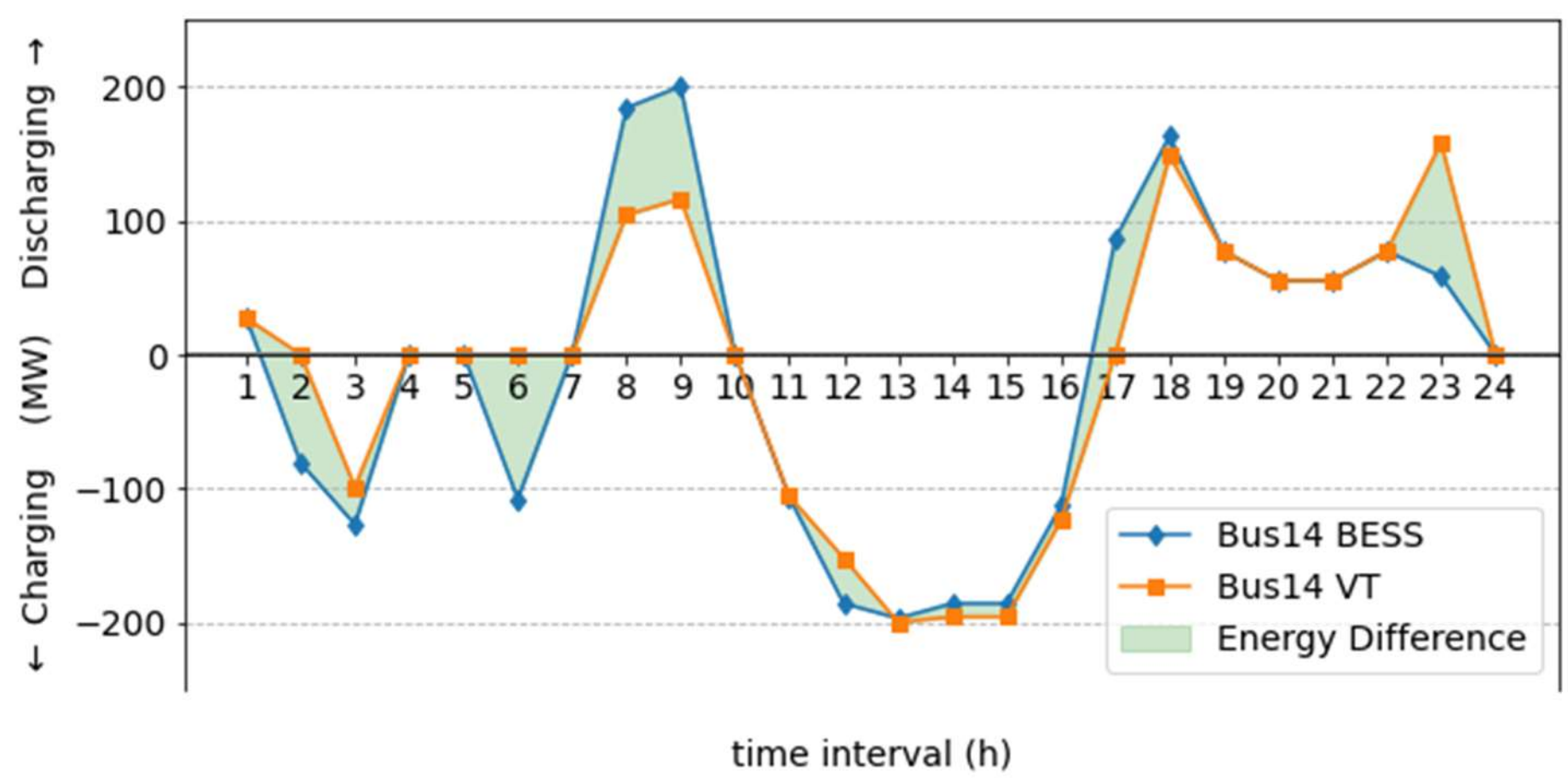

Figure 4. Charging and discharging profile of the battery at bus 14 under different CM strategies.

These two figures show results from both SCUC-BESS and SCUC-VT. The green area between the charging/discharging curves of the two CM schemes represents the difference in the energy being stored in the batteries when in the charging mode or in the energy that the batteries inject into the grid in the discharging mode.

The battery on bus 11 shows a higher charging/discharging rate in the SCUS-VT case than in the SCUS-BESS case. One possible explanation is that allowing only one battery in the charging mode will force the energy to be stored in a more concentrated manner. As a result, the battery on bus 11 in the VT case stores more energy before dawn, enabling it to provide more power between 11 a.m. and 4 p.m. when line 19 is congested, and line 29 gets stressed. In the BESS case, the average number of congested lines per hour rises to 0.42 from the SCUC base case's 0.38, mainly because generator 23 at bus 18 delivers more power during peak sun hours, causing already-stressed line 29 to become congested.

Whether with or without the VT operation constraint, the energy usages of the two BESS on each side of the transmission line are not balanced. With the solar and load profile of the test case, the battery on bus 14, close to the solar resources, is used heavier than the battery on bus 11.

*3.5 Market Analysis*

Other important metrics to evaluate VT compared with other CM schemes would be the energy market settlements, including the system operation cost and load payment, which reflects social welfare. These results are presented in Figure 5. The load payment of the VT scheme is lower than that of the standalone BESS scheme, while the load payment of the VT+NR scheme is higher than that of the BESS + NR schemes.

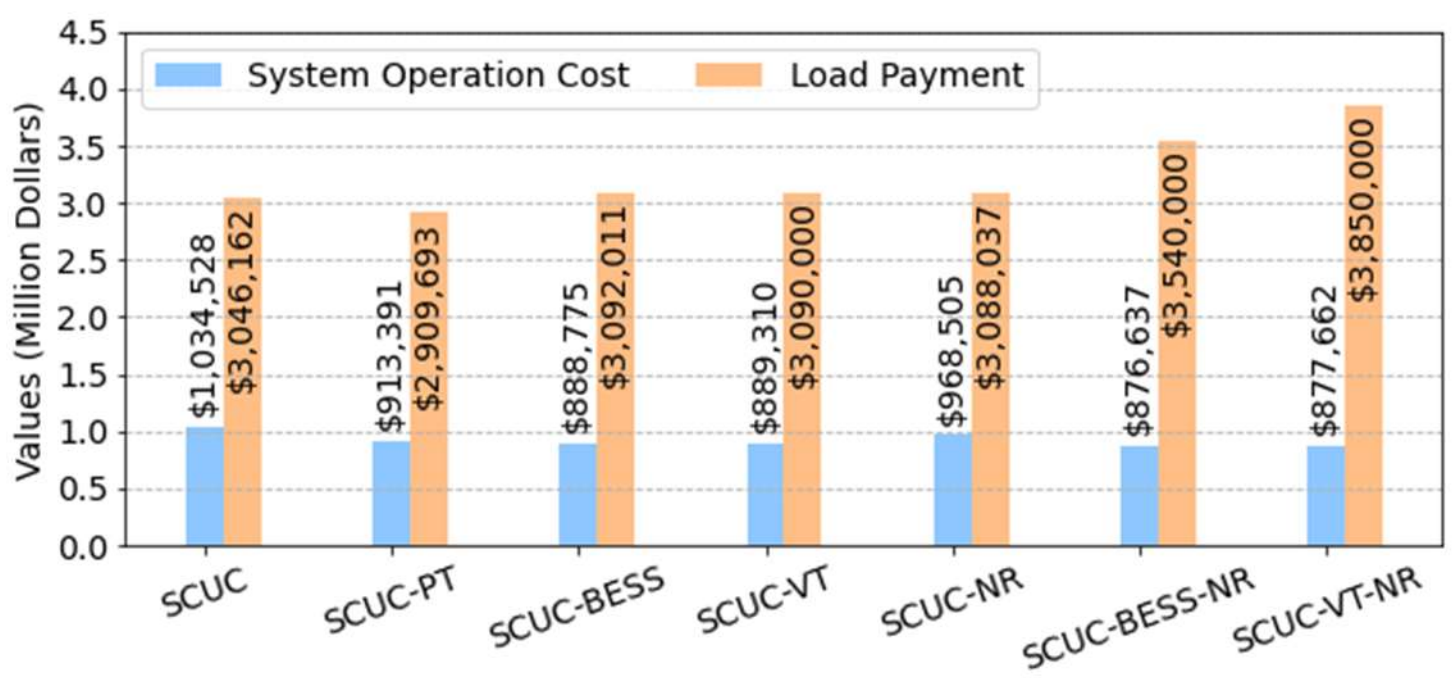

Figure 5. System operation costs and energy market settlement for different SCUC models under various CM schemes.

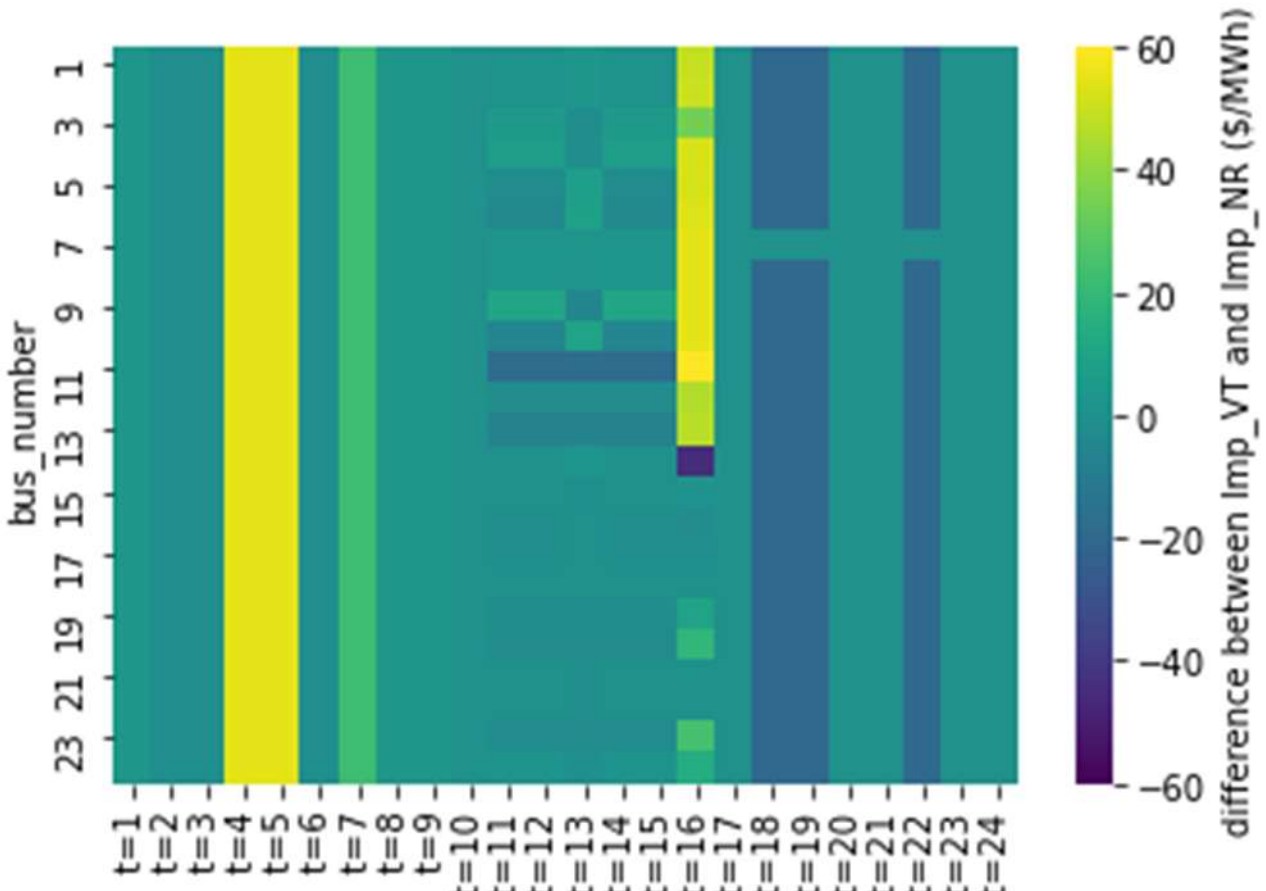

Figure 6. LMP difference between VT and NR cases.

Load payment is not necessarily correlated with system operation costs and congestion status. Although all six enhanced SCUC models with various CM schemes have less system operation cost, only the PT CM strategy leads to lower load payment than the SCUC benchmark. Although the VT CM scheme provides more congestion relief than the NR CM scheme, its load payment is higher than the NR CM scheme. More robust evidence can be observed from the results of SCUC-BESS-NR and SCUC-VT-NR, which lead to much more significant cost reduction but much higher load payment.

The load payment is related to the load profile because batteries re-shape the load profile through charging and discharging activities. Figure 6 illustrates the difference in LMP between the SCUC-VT model and the SCUC-NR model at each bus during each hour. It is observed that, compared with the SCUC-NR model, the SCUC-VT model significantly increases the LMP when the battery on bus 11 absorbs energy from the grid as an additional load, as shown in Figure 3, without enough low-cost generation online. The occasions are when the system is not congested, especially between 4 and 5 a.m. This explains why the load payment of SCUC-VT is higher even though SCUC-VT can lead to lower total cost.

*3.6 Sensitivity Analysis*

*3.6.1 Battery Size*

The size of the battery significantly impacts the battery charging/discharging decisions that affect the market settlement results. Figure 7 summarizes system operation costs and load payments for different BESS sizes, ranging from 100 to 400 MW, with an increment of 50 MW, assuming the same duration.

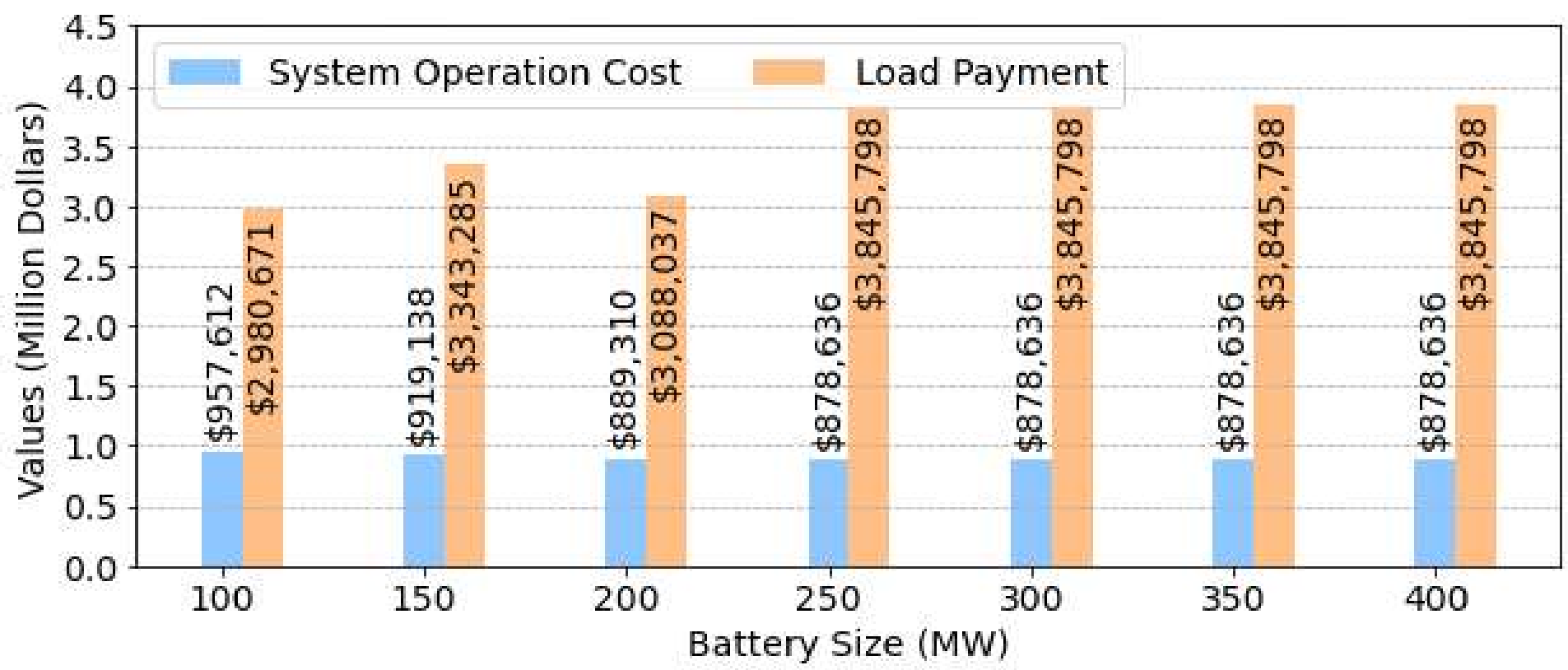

Figure 7 System operation cost and energy market settlement for different BESS sizes.

It is observed from Figure 7 that as the BESS size increases, the system operation cost reduces accordingly, which is expected. However, it is interesting to observe that the total cost remains the same after it drops to $878,636 when the BESS size increases to 250 MW; further increasing the BESS size will not provide any further benefits against network congestion. Similarly, the load payment does not change when BESS size reaches the exact turning point of 250 MW, precisely, a range of 200 MW - 250 MW. It is also interesting to observe that there is no fixed pattern regarding load payment change concerning BESS size change before BESS size hits this turning point.

The battery's size also affects the congestion status of transmission lines. As listed in Table 3The average number of congested lines per hour reduces when the BESSs' sizes on both sides of the line increase. Similar to the trend of operation cost, the average number of congested lines per hour does not change when the BESS size reaches the exact turning point of 250 MW.

Table 3 Line congestion status for different BESS sizes

| BESS size (MW) | 100 | 150 | 200 | 250 | 300 | 350 | 400 |
|---|---|---|---|---|---|---|---|
| Average No. of congested lines per hour | 0.38 | 0.25 | 0.25 | 0.17 | 0.17 | 0.17 | 0.17 |

Figure 8 shows the power flows on line 19 for two models: (i) SCUC benchmark and (ii) SCUC-PT with a new line parallel to line 19 and sharing the same parameters with line 19. For SCUC-PT, the total flow crossing the path of line 19 is slightly over 700 MW for the congestion hours from 12 pm to 3 pm, indicating that entirely relieving the congestion would require slightly over 200

MW additional transfer capacity, which aligns with the optimal BESS size of power capacity per Figure 7.

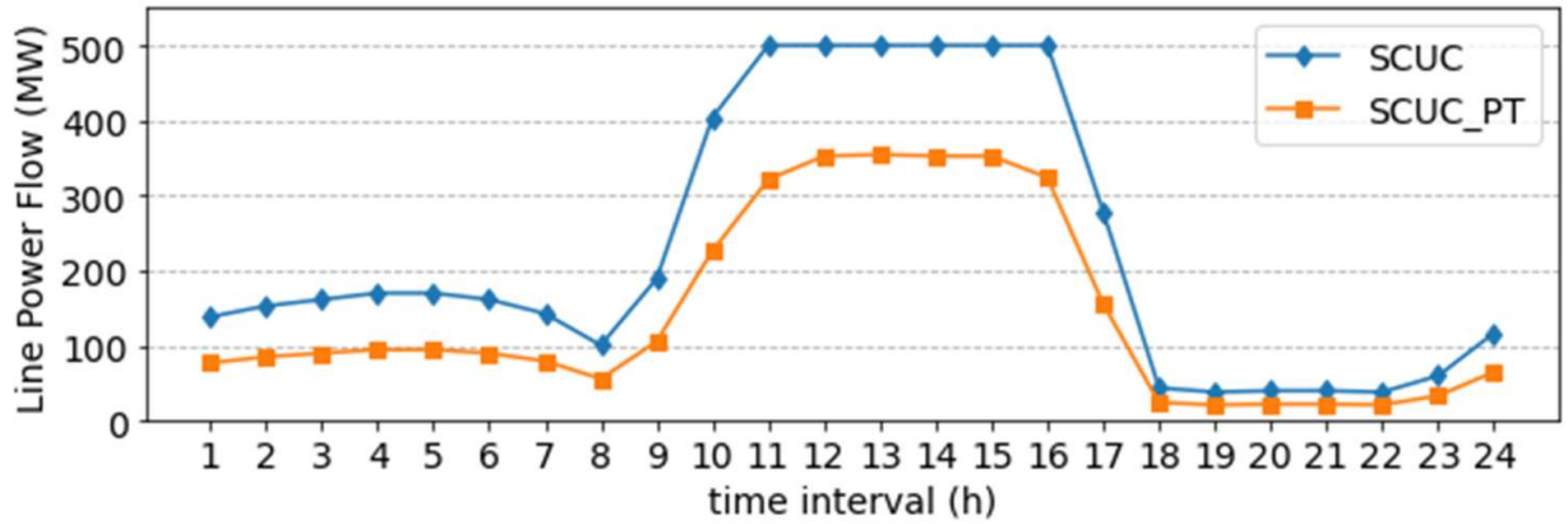

Figure 8 Power flows on line 19 from benchmark SCUC and SCUC-PT.

*3.6.2 Renewable Penetration Level*

The renewable power penetration level impacts the availability of cheap renewable resources, which may lead to lower operational costs. However, the network may get congested as the renewable penetration level increases. When there is no room to transfer a large amount of solar power, SCUC may need to find a feasible solution through curtailment. To evaluate the performance of different schemes under different penetration levels, we enabled solar and wind curtailment for the schemes to handle higher solar and wind penetration levels. A penalty in the objective function was introduced to minimize the curtailment. Equations (25)-(27) are the revised objective function. Equations (28)-(32) are the constraints to represent solar and wind generation curtailment.

Objective:

$$minimize \sum_{g\in G} \sum_{t\in T} \left( c_g P_{gt} + c_g^{NL} * u_{gt} + c_g^{SU} * v_{gt} \right) - C_s - C_w \tag{25}$$

Where:

$$C_s = \rho_s \sum_{s\in S} P_{st}^C \tag{26}$$

$$C_w = \rho_w \sum_{w\in W} P_{wt}^C \tag{27}$$

Constraints:

$$\begin{aligned}\sum_{g\in G(n)} P_{gt} + \sum_{k\in K(n-)} P_{kt} - \sum_{k\in K(n+)} P_{kt} \\ = \sum_{e\in ES(n)} \left(P_{et}^{c} - P_{et}^{d}\right) + d_{nt} - \sum_{s\in S(n)} \left(P_{st} - P_{st}^{C}\right) \\ - \sum_{w\in W(n)} \left(P_{wt} - P_{wt}^{C}\right), \forall n, t\end{aligned} \quad (28)$$

$$u_{st} \leq P_{st} BigM \quad (29)$$

$$P_{st}^{C} \leq P_{st} u_{st} \quad (30)$$

$$u_{wt} \leq P_{wt} BigM \quad (31)$$

$$P_{wt}^{C} \leq P_{wt} u_{wt} \quad (32)$$

This paper defines the renewable energy penetration level as the percentage of total renewable energy capacity to the peak load. Additional solar or wind generation has been added to buses 2, 4, 5, 8, 19, and 23.

Figure *9* System operation cost and energy market settlement for different solar power penetration levelsand shows system operation costs and load payments for different solar penetration levels, ranging from 50% to 100%, with an increment of 10%, assuming no change in the location and size of the batteries. Figure *10* shows system operation costs and load payments for different wind penetration levels on top of 40% solar penetration, ranging from 10% to 60%, with an increment of 10%, assuming no change in the location and size of the batteries.

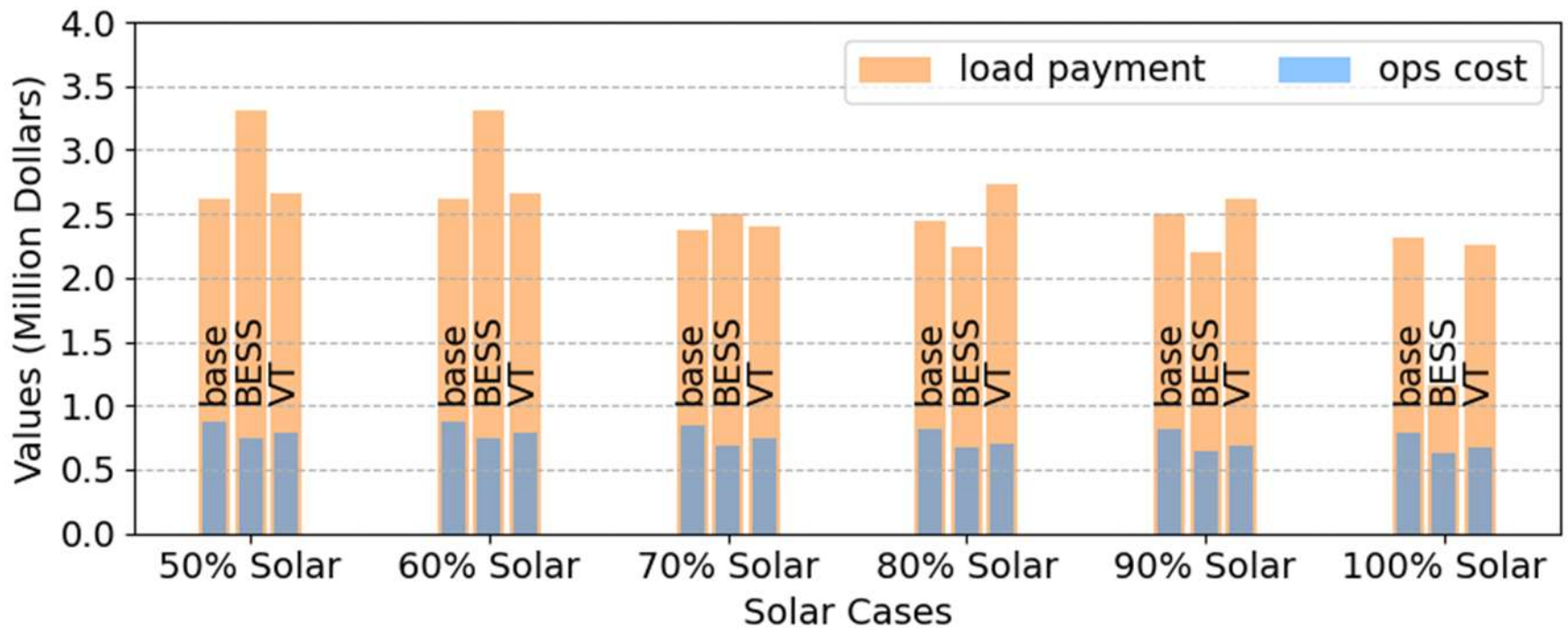

Figure 9 System operation cost and energy market settlement for different solar power penetration levels

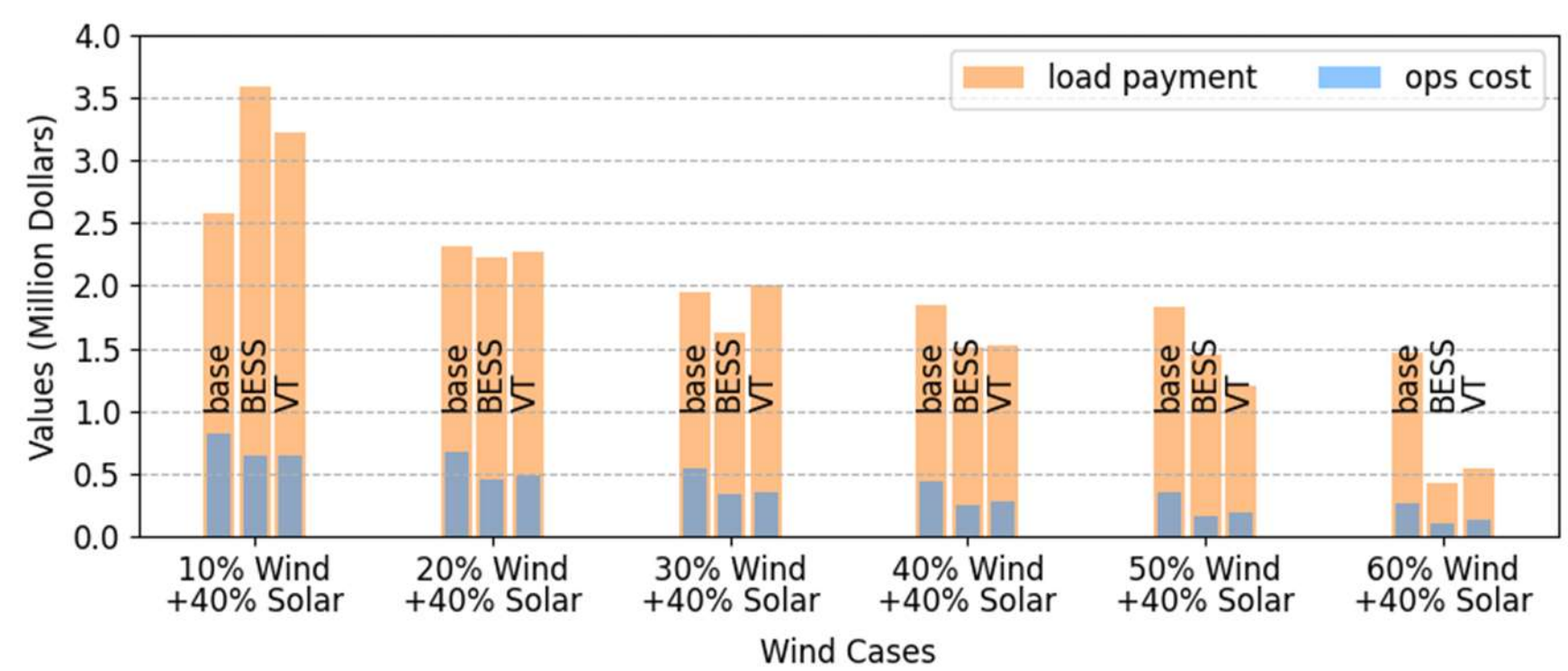

Figure 10 System operation cost and energy market settlement for different wind power penetration levels

The simulation results show that BESS and VT schemes can reduce operational costs under all circumstances. However, using BESS or VT schemes does not guarantee a lower load payment. The resource profile determines the load payment. There is a greater chance of observing a lower load payment with more wind penetration, which could provide cheap resources during battery charging mode. The VT scheme tends to provide a lower load payment than the BESS scheme under a lower renewable penetration level.

*Figure 11* and *Figure 12* compare the solar and wind curtailment in 24 hours under different solar and wind penetration levels. Renewable curtailment is needed when solar penetration reaches 70% or when solar and wind combined penetration reaches 90%. The TS scheme provides the best performance on curtailment reduction before renewable penetration reaches 100%. BESS and VT schemes could significantly reduce curtailment when renewable penetration reaches 100%.

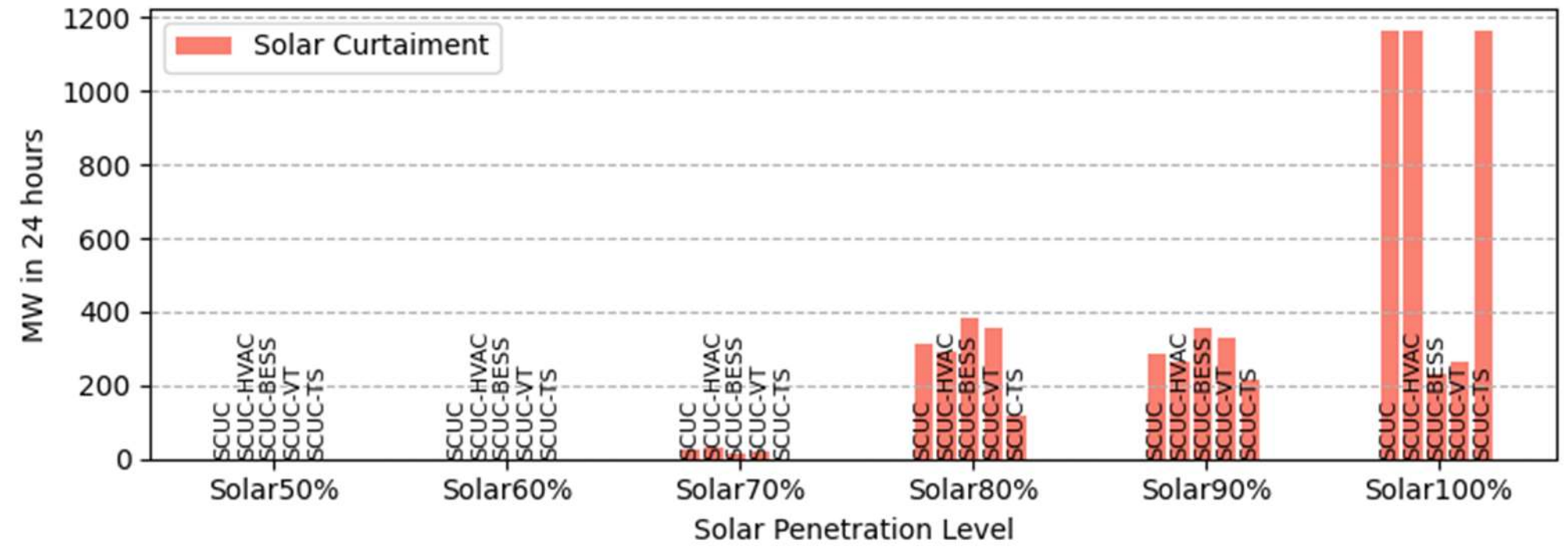

Figure 11 Solar Curtailment under different solar penetration levels

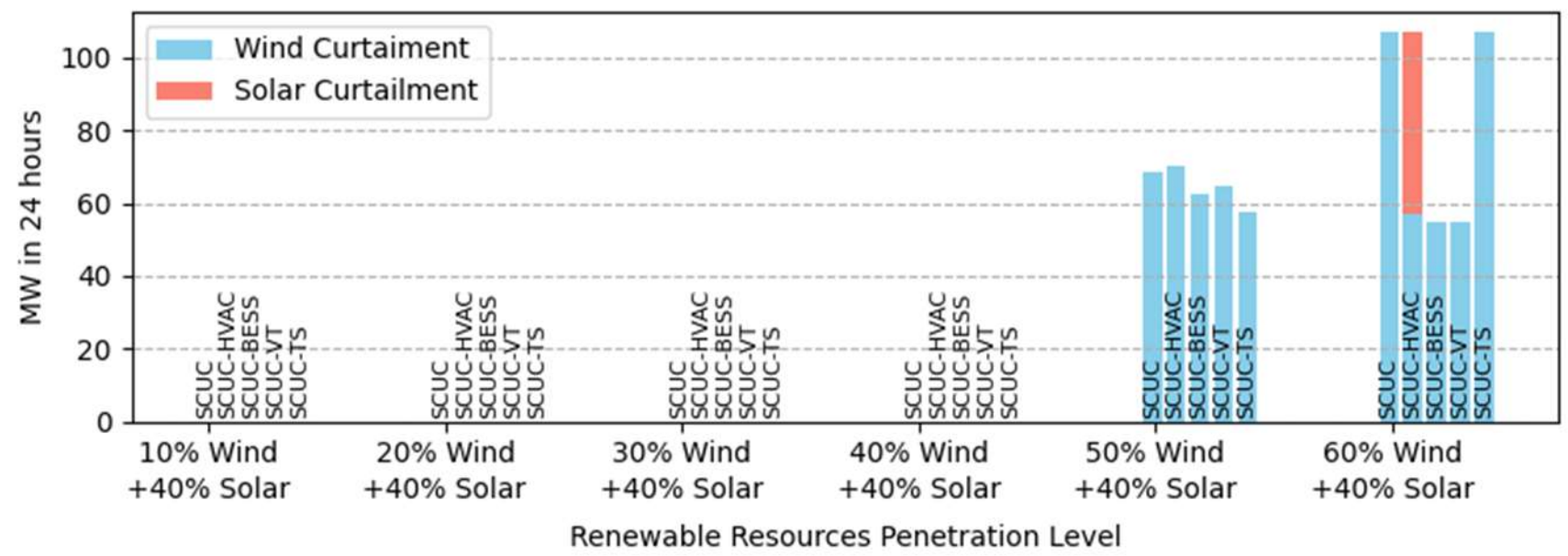

Figure 12 Solar and Wind curtailment under different wind penetration levels

Figure *13* and Figure *14* illustrate the line congestion status for different schemes under different solar and wind penetration levels. Adding a new physical transmission line generally provides the best results for relieving congestion. The VT scheme follows, beating the BESS scheme on congestion reduction under all circumstances.

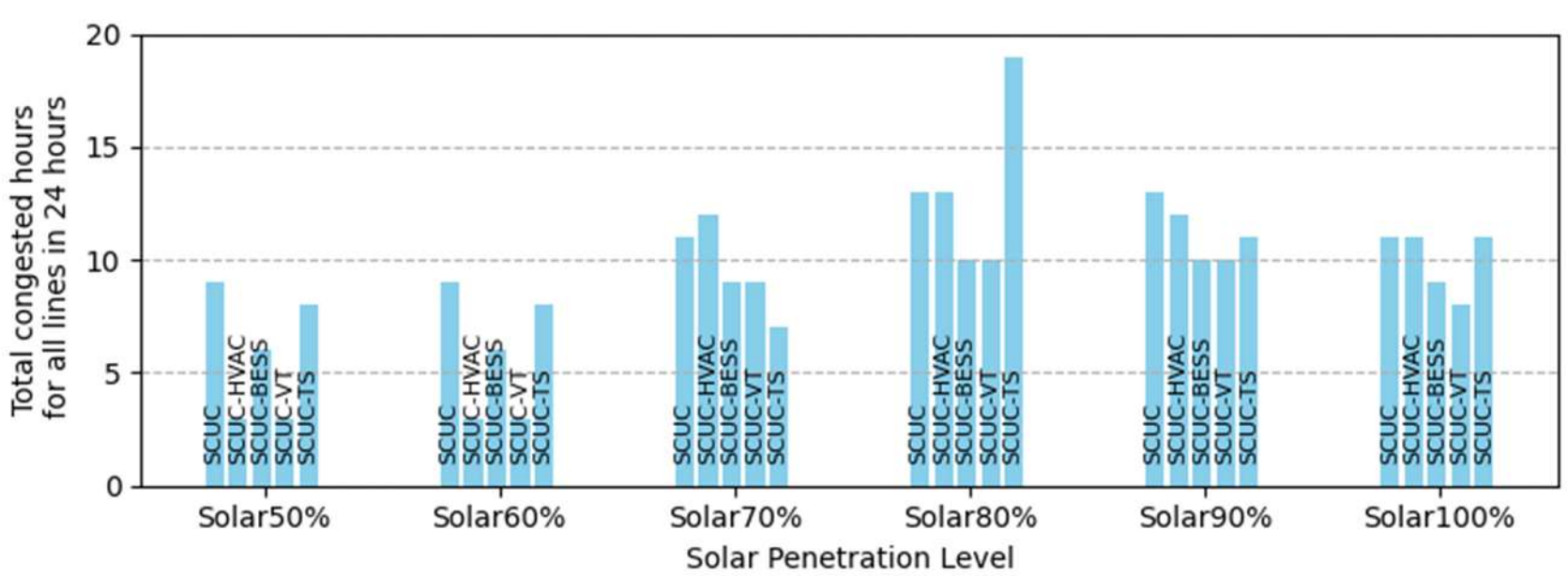

Figure 13 Total congested hours for all transmission lines in 24 hours under different solar penetration levels

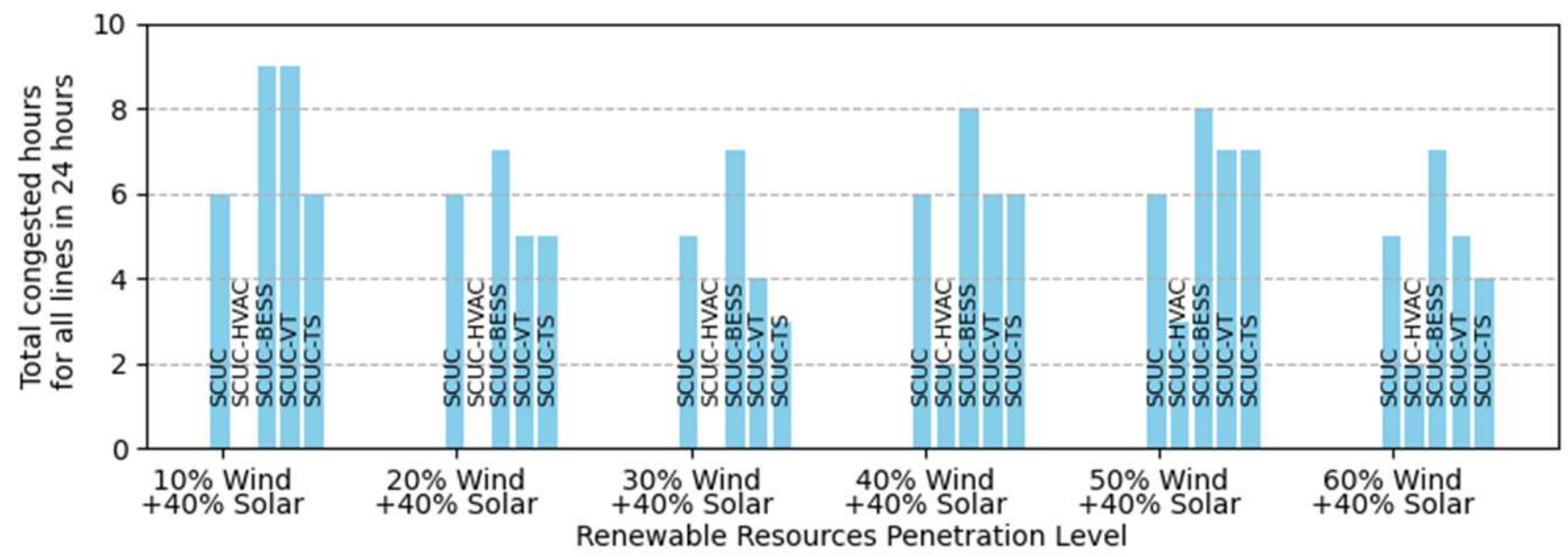

Figure 14 Total congested hours for all transmission lines in 24 hours under different wind penetration levels

*3.6.3 Load Profile*

The change in load profile throughout the year affects the power flow and the UC results. To evaluate the consistency of VT's performance in various load scenarios, we run the simulations using the weekday and weekend load profiles of the fall/spring season. The daily peak load of the fall/spring season is assumed to be 70% of the maximum load. The load profiled in the test cases uses fall/spring weekday and weekend data in [30] as the hourly peak load in percent of daily peak.

Table 4 summarizes the operation cost reduction and congestion conditions under different CM schemes. The study results show that BESS and VT schemes achieve the highest cost reduction, as demonstrated in the summer weekday load profile scenarios. BESS does not relieve congestion, while VT offers relief in the weekend scenario.

Table 4 Transmission Facility Performance Comparison for fall/spring season

| Model | Operation cost reduction | | Average No. of congested lines per hour | |
|---|---|---|---|---|
| Scenario | Fall/Spring Weekday | Fall/Spring Weekend | Fall/Spring Weekday | Fall/Spring Weekend |
| SCUC | 0.00% | 0.00% | 0.25 | 0.25 |
| SCUC-PT | 7.13% | 5.33% | 0 | 0 |
| SCUC-BESS | 21.19% | 19.32% | 0.25 | 0.25 |
| SCUC-VT | 18.77% | 16.86% | 0.25 | 0.208 |
| SCUC-NR | 6.27% | 4.76% | 0.208 | 0.208 |

## 4. Conclusions

This paper aims to comprehensively evaluate the performance and economic benefits of using BESS as VT compared to other non-wire solutions. The main contributions of this work are to 1) propose a practical VT working mode and associated model and 2) quantitatively benchmark the performance and economic benefits of VT against the physical line, standalone BESS, network reconfiguration, and the combination of schemes.

The study results demonstrate that BESS-based VT can help reduce the total grid operation cost, reduce load payment, and relieve network congestion. Compared with a new physical line or network reconfiguration strategy, BESS-based VT can achieve more significant cost reduction, shorter computational time, and less renewable curtailment. Compared with the standalone BESS scheme, BESS-based VT helps provide more system congestion relief and social benefits, usually a lower load payment. Careful BESS size design is required for VT to provide optimal performance. The battery size on each side of the critical line that may be congested in peak hours would affect the VT performance regarding congestion relief, cost reduction, and market clearing results. Combining VT with other system congestion-relieving methods like NR may significantly increase the optimization calculation time.

## 5. Future Work

This paper discusses the performance of VT in short-term operation based on the deterministic model and using the IEEE-24 test system. More evaluation on long-term planning and using real-world power systems remains to be explored.

Wind/solar forecast errors could introduce uncertainty in real-world applications. Stochastic unit commitment may be used to solve multiple system scenarios to produce more reliable solutions. The performance of VT, especially the computation time, needs to be evaluated in stochastic unit commitment.

In addition, further research is needed to (1) evaluate VT's behavior in power systems at different locations and (2) optimize the size of the BESS on each side of the critical line to achieve the optimum VT performance with the least cost.